\documentclass[aps,prl,twocolumn,superscriptaddress,showpacs]{revtex4-1}

\usepackage[utf8x]{inputenc} % should really switch over to XeTeX/LuaTeX ...
\usepackage[T1]{fontenc}
\usepackage[english,american]{babel}

\usepackage{lmodern}

\usepackage{textcomp,gensymb}
\usepackage{amssymb,amsmath}
\usepackage{units}
\usepackage{bm}
\usepackage{isotope}

\DeclareMathOperator{\elE}{\mathcal{E}}
\DeclareMathOperator{\elK}{\mathcal{K}}

\newcommand*{\dd}{\mathrm{d}\!} % differential d, replace with commath pkg?
\newcommand*{\ee}{\mathrm{e}} % Euler's e
\newcommand*{\prob}[1]{P(#1)}

\newcommand*{\kB}{k_\text{B}}

\newcommand*{\Sonehalf}{\mathrm{S}_{1/2}}
\newcommand*{\Ponehalf}{\mathrm{P}_{1/2}}
\newcommand*{\Dthreehalves}{\mathrm{D}_{3/2}}

\newcommand*{\Caforty}{\isotope[40]{Ca}}

\usepackage{graphicx}

\usepackage{hyperref}
\hypersetup{
  pdfstartview={FitH},
  pdftitle={},
  pdfauthor={Ian D. Leroux}}

\ifpdf
\fi

\begin{document}

\title{Pinning an Ion with an Intracavity Optical Lattice}

\newcommand*{\IFAAU}{
QUANTOP, Danish National Research Center for Quantum Optics, Department of Physics and Astronomy, Aarhus University, DK-8000 Aarhus C, Denmark}
\author{Rasmus B. Linnet} \affiliation{\IFAAU}
\author{Ian D. Leroux} \affiliation{\IFAAU}
\author{Mathieu Marciante} \affiliation{Aix-Marseille Université, CNRS, PIIM, UMR 7345, Centre de Saint Jérôme,
Case C21, 13397 Marseille Cedex 20, France}
\author{Aurélien Dantan} \affiliation{\IFAAU}
\author{Michael Drewsen} \email{drewsen@phys.au.dk} \affiliation{\IFAAU}
\date{\today}


\begin{abstract}
We report one-dimensional pinning of a single ion by an optical lattice.
A standing-wave cavity produces the lattice potential along the rf-field-free axis of a linear Paul trap.
The ion's localization is detected by measuring its fluorescence when excited by standing-wave fields with the same period,
but different spatial phases.
The experiments agree with an analytical model of the localization process,
which we test against numerical simulations.
For the best localization achieved,
the ion's average coupling to the cavity field is enhanced from 50\% to 81(3)\% of its maximum possible value,
and we infer that the ion is bound in a lattice well with over 97\% probability.
\end{abstract}
\maketitle

The fields of ultracold trapped neutral atoms and ions generally rely on unrelated trapping technologies.
Ions can be trapped using Lorentz forces,
including pure Coulomb forces.
Such traps can be very deep ($\unit[\sim 10^5]{K}$),
but have typical length scales of at least $\unit[0.1]{mm}$ because parasitic surface charges disturb the ion's motion if electrodes are too close to it.
Neutral atom traps require position-dependent perturbation of internal energy levels,
generally induced by optical or magnetic fields.
The resulting potentials are shallower
(typically a few tens of $\unit{mK}$ or less),
but can have optical-wavelength-scale structure~\cite{Grimm2000,Schmied2010}.
Confining ions using such finely tailored potentials would be valuable for quantum simulations of many-body physics with ion crystals~\cite{Porras2004,Friedenauer2008,Schmied2008,Kim2010,Britton2012,Schneider2012:review},
for studies of the Coulomb-Frenkel-Kontorova model for friction~\cite{Garcia-Mata2007,Benassi2011,Pruttivarasin2011},
for studies of dynamical localization~\cite{ElGhafar1997},
for studies of particles in potentials with significant quantum fluctuations of their own~\cite{Bushev2004,Cormick2012},
and for optimized spatial phase matching of ion crystals to optical modes in cavity QED experiments~\cite{Herskind2009:coupling,Albert2011}.

Confining ions via internal state manipulation is challenging because the achievable forces,
though just as strong in ions as in atoms~\cite{Cormick2011},
are weak compared to typical Coulomb forces in ion traps~\cite{Schneider2012:fields}.
Anomalously slow ion diffusion through a polarization lattice has been observed~\cite{Katori1997},
and purely optical ion trapping in a Gaussian beam of $\unit[7]{\micro m}$ waist radius has been demonstrated~\cite{Schneider2010},
but ion confinement in an optical-wavelength-scale potential has so far remained an open problem.

Here we demonstrate the pinning of an ion,
held in a linear Paul trap,
by the one-dimensional lattice potential generated by the standing-wave field in an optical resonator.
As our imaging system cannot resolve lattice sites,
we observe the lattice's effects using the ion's fluorescence in structured driving fields~\cite{Guthoehrlein2001,Mundt2002,Eschner2003,Walther2012}.
One such field is the lattice itself,
whose scattering rate is a measure of the ion's average potential energy.
We also employ a near-detuned standing-wave probe field,
verifying that its scattering is suppressed when probe intensity minima align with lattice wells.
We present a classical analytical model which quantitatively agrees with our observations.

\begin{figure}
  \centering
  \includegraphics{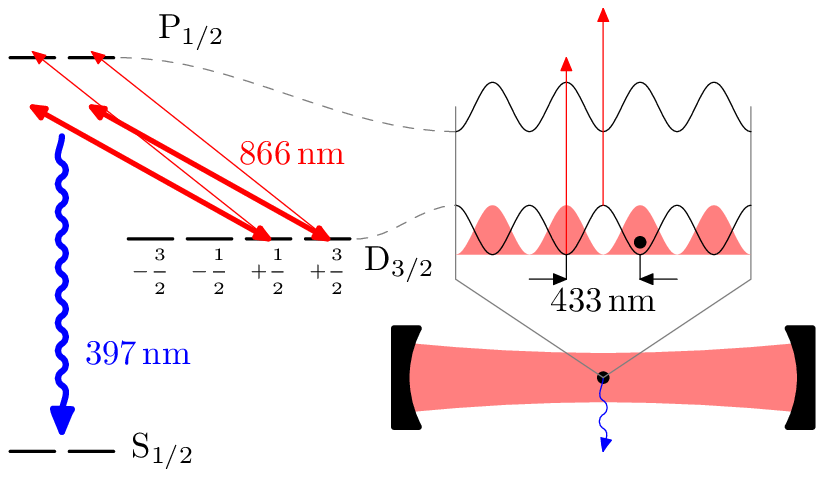} 
  \caption{Left: Relevant levels in $\Caforty^+$.
    An ion in the metastable $\Dthreehalves$ state is pinned by a far-detuned lattice field (thick arrows).
    A near-detuned probe field (thin arrows) tests the ion's position distribution.
    The dominant scattering process is inelastic scattering to the $\Sonehalf$ dark state,
    producing observable $\unit[397]{nm}$ fluorescence (wavy arrow).
    Right: Position-dependent $\Dthreehalves$ and $\Ponehalf$ energy levels of the ion on the cavity axis.
    Intracavity standing waves can be either in-phase or out-of-phase (shading) with the periodic potential for $\Dthreehalves$,
    suppressing or enhancing fluorescence for a pinned ion.
    Lattice Stark shifts also affect the probe detuning (vertical arrows).}
  \label{fig:setup}
\end{figure}

The apparatus is described in Refs.~\cite{Herskind2008,Herskind2009:loads},
and its essential elements are sketched in Fig.~\ref{fig:setup}.
A single $\Caforty^+$ ion is held in a linear Paul trap operating at a $\unit[4.0]{MHz}$ drive frequency,
with radial and axial trapping frequencies of $\unit[377]{kHz}$ and $\unit[97]{kHz}$, respectively.
Every experimental cycle begins with $\unit[28]{\micro s}$ of Doppler cooling on the $\Sonehalf\rightarrow\Ponehalf$ transition,
with a repumper on $\Dthreehalves\rightarrow\Ponehalf$,
preparing the ion in an approximately thermal motional state~\cite{Metcalf1999}.
The ion is then pumped into $\Dthreehalves$ by switching off the repumper for $\unit[14]{\micro s}$.
Finally, 
the cooling light is extinguished and a $\unit[433]{nm}$-period optical lattice is generated by driving an $\unit[11.7]{mm}$-long near-confocal cavity of finesse $3000$ and waist radius $\unit[37]{\micro m}$,
whose optical axis coincides with the nodal line of the Paul trap~\cite{Herskind2009:loads}.
The lattice is $\sigma^-$-polarized relative to a $\sim\unit[1]{G}$ bias field along the cavity axis and is detuned from the $\Dthreehalves\leftrightarrow\Ponehalf$ transition by $\sim\unit[0.2]{THz}$.
The optical potential is ramped up in $\unit[2]{\micro s}$,
held high for $\unit[5]{\micro s}$ while we observe the ion's fluorescence,
and then switched off before the cycle repeats.

As we discuss below,
the fluorescence process used to observe the ion also leads to loss of the ion from the internal states which feel the lattice potential.
For long hold times,
the probability of ion loss approaches unity,
causing the fluorescence signal to saturate to an uninteresting constant.
The $\unit[5]{\micro s}$ hold time is a compromise choice,
long enough to yield a detectable fluorescence signal but shorter than the ion's lifetime in the lattice.
This hold time can be extended in future experiments by reducing the lattice scattering rate,
either by increasing the lattice laser detuning or,
in the case of a blue-detuned lattice whose scattering rate scales with the ion's potential energy,
by cooling the ion further.

Certain features of this experiment allow us to describe the lattice's effect with a simple analytical model,
enabling quantitative data analysis.
First, note that the lattice drives an open transition from the metastable $\Dthreehalves$ state.
96\% of photon scattering events transfer the ion to either $\Sonehalf$ or the substates $m=-1/2, -3/2$ of $\Dthreehalves$,
where it plays no further role in the experiment.
Hence, an observable ion has not been subject to significant dissipative radiation-pressure forces while in the lattice,
and its dynamics are those of a classical particle in a conservative potential.

A second simplification is that the Paul trap is loose enough that the ion position distribution extends over many lattice sites.
It follows that the change in the harmonic trap potential within a single site is small compared to the thermal energy,
and hence the potential in each site has approximately the same shape.
Since the momentum distribution in the initial thermal state is also the same everywhere,
the ion's motion can be adequately described using a single site with periodic boundary conditions.

Finally, the $\unit[2]{\micro s}$ lattice ramp-up is slow compared to typical oscillation frequencies in a site ($\gtrsim2\pi\times\unit[1]{MHz}$),
such that the action of the ion's trajectory is conserved (see Ref.~\cite{Wells2007}, for instance)\footnote{
  Strictly speaking, this argument cannot be applied to trajectories near the unstable equilibrium point at the top of a lattice barrier, for which the oscillation period diverges.
  The problematic region of phase space is small enough not to affect significantly the accuracy of our results.}.
Knowing the action's distribution in the initial thermal state and the action as a function of energy in the sinusoidal lattice potential,
we predict the energy distribution in the lattice
\begin{equation}
  \prob{E} \dd E =  \frac{\ee^{-s(E)^2 / 4 \kB T_0}}{\sqrt{\pi \kB T_0}}\tau(E) \dd E,
  \label{eq:ssaEpdf}
\end{equation}
where $E$ is the energy (measured from the local trap potential),
$\kB T_0$ is the initial thermal energy,
$\tau$ is the period of the trajectory normalized to that of small oscillations in the lattice
\begin{equation}
  \tau(E) = \frac{2}{\pi}\times
  \begin{cases}
    \elK(E) & E \le 1 \\
    \sqrt{E^{-1}} \elK(E^{-1}) & E > 1,
  \end{cases}
  \label{eq:period}
\end{equation}
and $s$ is the corresponding dimensionless action
\begin{equation}
  s(E) = \frac{4}{\pi}\times
  \begin{cases}
    \elE(E) - (1 - E) \elK(E) & E \le 1 \\
    \sqrt{E} \elE(E^{-1}) & E > 1,
  \end{cases}
\end{equation}
with $\elK$ and $\elE$ being the complete elliptic integrals of the first and second kind, respectively.
All energies are expressed in units of the lattice depth.

\begin{figure}
  \centering
  \includegraphics{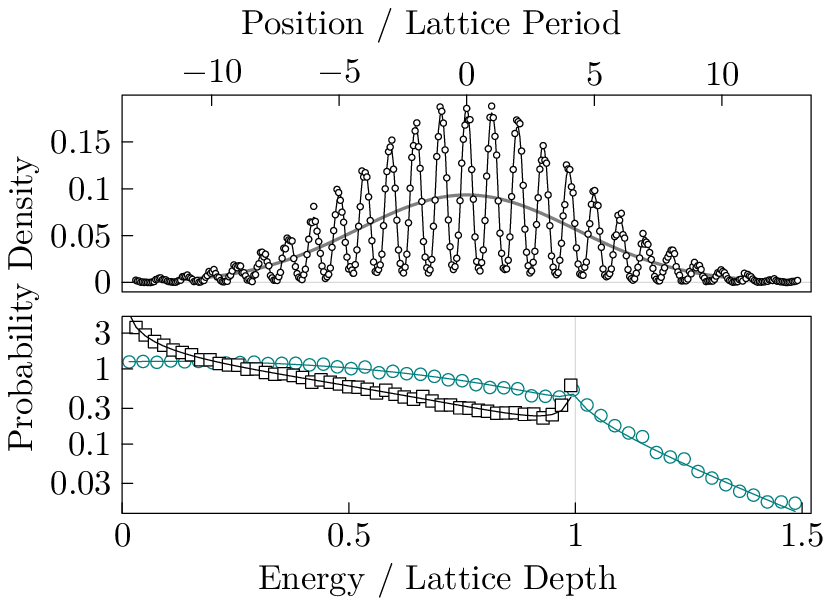}
  \caption{Comparisons of model predictions (lines) to molecular dynamics simulations (symbols).
    Upper graph: Initial thermal position distribution (gray Gaussian) and final position distribution in the lattice (black line and symbols).
    Lower graph: Distributions of total energy (teal line, circles) and lattice potential (black solid line, squares), 
    measured relative to the local trap potential.
    Note logarithmic vertical scale.
    The simulations use typical experimental parameters,
    with a $\unit[5]{mK}$ initial temperature and a $\unit[24]{mK}$ final lattice depth.}
  \label{fig:modelling}
\end{figure}

This total energy distribution and the associated lattice potential distribution are shown together with the position distribution in Fig.~\ref{fig:modelling} for typical experimental parameters,
along with histograms obtained from molecular dynamics (MD) simulations of our experiment.
The model position distribution (upper graph) illustrates the nature of the localization:
the ion's global position uncertainty remains as large as in the initial thermal state (grey Gaussian),
but in any given experimental realization the ion has a high probability of being pinned in a lattice well aligned with the cavity field.
The vertical line in the lower plot indicates the top of the lattice.
The portion of the total energy distribution (teal line, circles) to the left of this line is pinned in a lattice well.
For all three distributions,
there is excellent agreement between the full MD simulations,
including the effects of finite lattice ramp-up time and of the harmonic trap potential,
 and the analytical model.
This tests our assumptions and confirms that a single-site model with adiabatic lattice ramp-up describes the essential physics of the experiment.

Experimentally, we can observe neither the full energy distribution nor the interesting subwavelength features of the position distribution directly.
To detect the lattice's effect on the ion,
we first observe the scattering of lattice light,
which varies with the ion's potential energy.
Inelastic scattering to the $\Sonehalf$ level is the dominant scattering process (93\% probability),
and it produces a background-free $\unit[397]{nm}$ fluorescence signal which can be observed with shot-noise-limited resolution by an image-intensified CCD camera.
By comparing this fluorescence signal to one obtained when the ion is deterministically depumped to $\Sonehalf$ with a resonant laser pulse,
we obtain a calibrated measure of the ion's probability to scatter a lattice photon.

\begin{figure}
  \centering
  \includegraphics{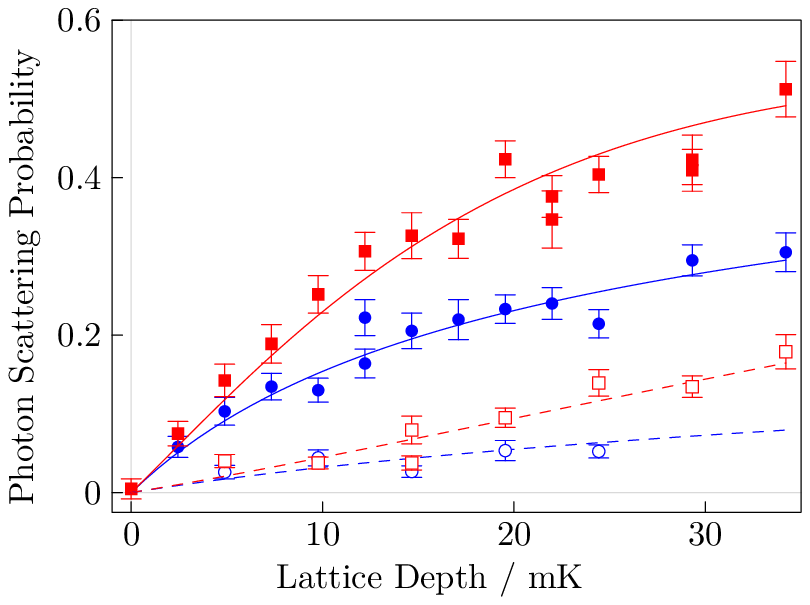}
  \caption{Solid symbols: probability to scatter a lattice photon as a function of lattice depth for red- and blue-detuned lattices (squares and circles respectively).
    Solid curves: one-free-parameter model fit.
    Open symbols: probability to scatter a photon during lattice ramp-up.
    Dashed lines: model prediction.}
  \label{fig:lattice-scattering}
\end{figure}

To obtain an unambiguous localization signal,
we compare the photon scattering probability for red- and blue-detuned lattices.
These generate identical potentials with the same maximum scattering rate,
but a red-detuned lattice pins the ion at lattice antinodes,
increasing scattering beyond the delocalized average,
while a blue-detuned lattice pins the ion at the nodes,
where scattering is suppressed.
Differences in scattering between the two scenarios are a signature of ion localization in the lattice potential.
Figure~\ref{fig:lattice-scattering} shows the probability to scatter a lattice photon in the $\unit[5]{\micro s}$ after ramp-up for detunings of $\pm\unit[0.19]{THz}$ (solid red squares and blue circles).
The regular spacing of the cavity modes ensures that the absolute detunings of the two lattices are equal to within one part in $10^4$.
The observed scattering is systematically higher for the red-detuned lattice,
indicating that the ion spends the majority of its time in the lower half of the lattice potential.

The initial optical pumping into $\Dthreehalves$ does not polarize the ion,
which has equal probability to be in each of the four Zeeman substates.
An ion in $m=-3/2$ or $-1/2$ does not couple to the lattice field and is ignored.
Of the remaining substates,
$m=+3/2$ has the strongest transition dipole moment and sees a lattice three times deeper than the one seen by $m=+1/2$.
We plot raw data obtained by observing both bright levels,
but report lattice depths and localization results for $m=+3/2$.
In future experiments with several ions,
optical pumping into $m=+3/2$ will ensure that all ions see the stronger lattice~\cite{Herskind2009:coupling}.

The solid curves in Fig.~\ref{fig:lattice-scattering} are model-predicted fluorescence signals,
computed from the known intensity profile of the intracavity fields and the model-supplied ion position distribution for each Zeeman component.
The model accounts for ion loss during lattice ramp-up,
which we have also measured as an independent cross-check
(Fig.~\ref{fig:lattice-scattering}, open symbols and dashed lines).
We find that this simple model provides a good quantitative description of the observed scattering probabilities.
The pair of curves are fit with the initial temperature as the only free parameter.
We obtain a temperature of $\unit[5.1(6)]{mK}$,
ten times the Doppler limit and consistent with the ion's observed position spread in the Paul trap.

We have verified that red- and blue-detuned lattices produce indistinguishable scattering signals if the ion is forcibly delocalized by driving it with an axial rf electric field.
Indeed, the localization experiment requires careful minimization of stray axial rf fields produced by the Paul trap~\cite{Berkeland1998}.
Fields driving only a few nanometers of motion in a free ion suffice to destroy the localization signal in both laboratory and numerical experiments,
presumably because the ion's response is resonantly enhanced near the $\unit[28]{mK}$ lattice depth for which the oscillation frequency in a well matches the trap drive frequency.
Suppressing trap-driven rf motion is a difficulty particular to Paul traps,
but it can be overcome by turning off the rf drive while the ion is in the lattice~\cite{Schneider2010,Schneider2012:fields}.

\begin{figure}
  \centering
  \includegraphics{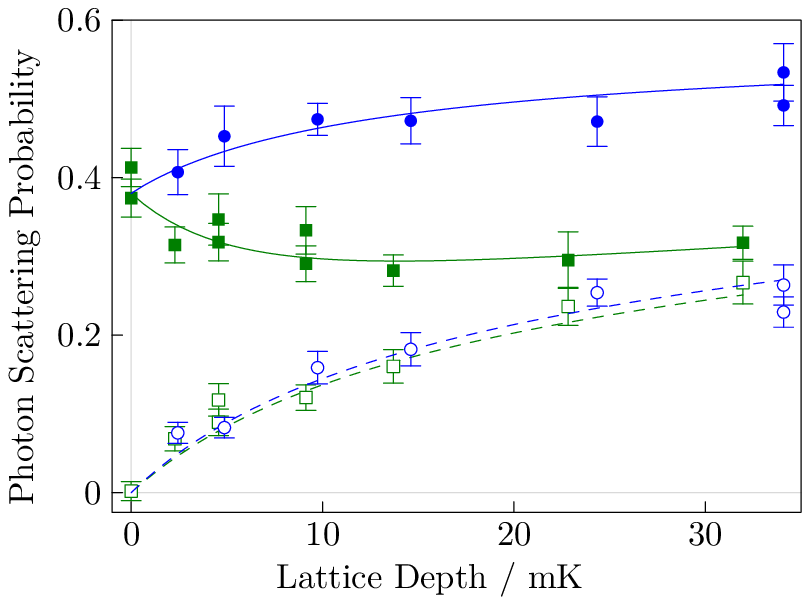}
  \caption{Solid symbols: combined scattering probability for blue-detuned probe and lattice fields separated by 15 (blue circles) and 16 (green squares) cavity free spectral ranges.
    Open symbols: scattering probability from the lattice alone.
    Curves: two-free-parameter fit.}
  \label{fig:probe-scattering}
\end{figure}

As a second measure of the ion's position distribution,
we excite the pinned ion with a near-resonant probe field and observe the resulting increase in photon scattering.
The boundary conditions imposed by the cavity mirrors ensure that,
at the ion's location in the center of the cavity,
the standing wave of the probe field overlaps with that of the lattice when they are separated in frequency by an even number of cavity free spectral ranges.
In this case,
localization of the ion enhances or suppresses scattering from both fields in the same way.
When the fields are separated by an odd number of free spectral ranges,
the nodes of the probe field align with the antinodes of the lattice and suppression of one scattering rate accompanies enhancement of the other.
The probe scattering has an additional position dependence due to the lattice-induced Stark shift,
since the probe detuning from the $\Dthreehalves\rightarrow\Ponehalf$ transition changes by twice the local lattice potential (Fig.~\ref{fig:setup}).
We detune the probe by $\unit[0.65]{GHz}$ to be less sensitive to this shift.
The observed scattering probabilities for blue lattice detunings of $\unit[0.20]{THz}$
(16 free spectral ranges from the probe)
and $\unit[0.19]{THz}$
(15 free spectral ranges from the probe)
are shown in Fig.~\ref{fig:probe-scattering}
(green squares and blue circles respectively).
The 6.7\% change in lattice detuning triples the probe field's contribution to the scattering signal,
as the ion goes from being pinned near probe nodes to being pinned near probe antinodes. 
This is visible in Fig.~\ref{fig:probe-scattering} as an increased separation between the signal with the probe on (solid symbols) and the lattice-only signal (open symbols).
Again, model predictions agree with the observed scattering rates.
We fit all four curves simultaneously with two free parameters:
the initial ion temperature [$\unit[3.9(3)]{mK}$] and the probe power.
The probe power obtained in the fit is consistent with that independently measured in the experiment.

For the strongest localization ($\unit[3.9]{mK}$ initial temperature, $\unit[34]{mK}$ final lattice depth),
the ion's average coupling to an appropriate cavity mode increases from 50\% of its maximum value for a delocalized ion to 81(3)\% when the ion is pinned.
This finding is independent of the details of our model,
in that the scattering rates we measure are directly proportional to cavity coupling,
and we need only correct for loss during lattice ramp-up,
which we have independently measured,
and for the signal from the $m=+1/2$ component.
To infer a more detailed distribution,
we interpret the scattering data using the single-site model,
supported by its good agreement with numerical simulations and with probe and lattice scattering data over a range of lattice depths.
The inferred energy distribution shows that the ion is captured in a single lattice well with over 97\% probability.

We have demonstrated pinning of an ion by an optical-wavelength-scale potential.
We propose to apply the same techniques to ion strings,
introducing a competition between the lattice potential and the Coulomb repulsion between ions,
leading eventually to the experimental study of the Coulomb-Frenkel-Kontorova model for ion strings and lattices with incommensurate spacing~\cite{Garcia-Mata2007,Benassi2011,Pruttivarasin2011},
the study of structural phases of ion crystals in quantum potentials~\cite{Cormick2012},
structural control of large Wigner crystals,
and the pinning of ions to cavity field antinodes to maximize cooperativity in cavity QED experiments~\cite{Herskind2009:coupling,Albert2011}.

We note that closely related results are reported in Ref.~\cite{Enderlein2012}.

We acknowledge helpful discussions with P. Horak.
M. Marciante was supported by the CNRS in the frame of a PICS.
This work was supported by the European Commission (STREP PICC and ITN CCQED) and the Carlsberg Foundation.

\bibliography{ions}

\end{document}